\documentclass[12pt,preprint]{aastex}

\slugcomment{Accepted for ApJ}
\shorttitle{Additional Ultracool White Dwarfs in SDSS}
\shortauthors{Harris et al.}

\begin{document}

\title{Additional Ultracool White Dwarfs Found in the Sloan Digital Sky Survey}

\author{Hugh C. Harris\altaffilmark{1},
Evalyn Gates\altaffilmark{2},
Geza Gyuk\altaffilmark{2,3},
Mark Subbarao\altaffilmark{2,3},
Scott F. Anderson\altaffilmark{4},
Patrick B. Hall\altaffilmark{5},
Jeffrey A. Munn\altaffilmark{1},
James Liebert\altaffilmark{6},
Gillian R. Knapp\altaffilmark{7},
D. Bizyaev\altaffilmark{8},
E. Malanushenko\altaffilmark{8},
V. Malanushenko\altaffilmark{8},
K. Pan\altaffilmark{8},
Donald P. Schneider\altaffilmark{9},
J. Allyn Smith\altaffilmark{10} }

\altaffiltext{1}{U.S. Naval Observatory, Flagstaff Station,
 Flagstaff, AZ 86001; hch@nofs.navy.mil}
 \altaffiltext{2}{The Univ. of Chicago, Department of Astronomy \& Astrophysics, 5640 S. Ellis Ave., Chicago, IL 60637}
\altaffiltext{3}{Adler Planetarium and Astronomy Museum, Chicago, IL 60605}
\altaffiltext{4}{Department of Astronomy, University of Washington,
   Seattle, WA 98195}
\altaffiltext{5}{York University, North York, ON}
\altaffiltext{6}{Steward Observatory, Univ. of Arizona, Tucson, AZ 85721}
\altaffiltext{7}{Princeton Univ. Observatory, Peyton Hall, Princeton, NJ 08544}
\altaffiltext{8}{Apache Point Observatory, PO Box 59, Sunspot, NM 88349-0059}
\altaffiltext{9}{The Pennsylvania State Univ., University Park, PA 16802}
\altaffiltext{10}{Austin Peay State Univ., Department of Physics and
   Astronomy, Clarksville, TN 37044}

\begin{abstract}
We identify seven new ultracool white dwarfs discovered in the Sloan Digital
Sky Survey (SDSS).  The SDSS photometry, spectra, and proper motions are
presented, and additional {\it BVRI} data are given for these and other
previously discovered ultracool white dwarfs.  The observed colors span a
remarkably wide range, qualitatively similar to colors predicted by models
for very cool white dwarfs.  One of the new stars (SDSS J1251+44) exhibits
strong collision-induced absorption (CIA) in its spectra, while the spectra
and colors of the other six are consistent with mild CIA.  Another of the
new discoveries (SDSS J2239+00A) is part of a binary system -- its companion
is also a cool white dwarf, and other data indicate that the companion
exhibits an infrared flux deficiency, making this the first binary system
composed of two CIA white dwarfs.  A third discovery (SDSS J0310$-$00)
has weak Balmer emission lines.  The proper motions of all seven stars
are consistent with membership in the disk or thick disk.
\end{abstract}


\section{Introduction}

As white dwarfs cool below an effective temperature of roughly 4000~K,
the density of gas in the photosphere increases to a point where
models of the atmosphere must include effects not normally important
in standard stellar models.  Absorption by molecular hydrogen,
referred to as collision-induced absorption (CIA), becomes dominant at
near-infrared wavelengths in the high-density gas.  It affects the
radiative energy transport in atmospheres of brown dwarfs, but it
can severely modify the emergent broadband colors of cool white dwarfs
(Bergeron, Saumon, Wesemael 1995a; Saumon \& Jacobson 1999; Hansen 1999;
Gates et al 2004).  Calculations of CIA (J{\o}rgensen et al. 2000;
Borysow et al. 2001) to be used as an opacity source in stellar models
have become more realistic, but further improvements can and are being
made.  Other opacity effects (Kowalski \& Saumon 2006) and non-ideal
gas treatment (Kowalski 2006, and references therein) also become
important, and challenge current models for both white dwarfs
and cool brown dwarfs.

Because most stars evolve into white dwarfs, the cool white dwarfs
found now in the solar neighborhood provide a record of star formation
in the Galactic halo, the thick disk, and the early history of the
thin disk.  Utilizing white dwarfs to extract star formation histories
requires separating these populations of white dwarfs; for example,
Reid (2005) describes a procedure to accomplish this separation,
but it is necessarily based on some simplifying, model-dependent
assumptions such as the mass distribution of white dwarfs in each
population.  For applying white dwarfs to study the Galactic halo
and thick disk and the age of the thin disk, knowing the origin and
evolution of the coolest white dwarfs is required.

Only recently have any white dwarfs been identified that exhibit CIA:
these include LHS~1126 (Bergeron et al. 1994), WD0346 (Hambly et al. 1997),
LHS~3250 (Harris et al. 1999), LHS~1402 (Oppenheimer et al. 2001),
SDSS J1337+00 (Harris et al. 2001a), five new SDSS discoveries (Gates
et al. 2004), GD~392B (Farihi 2004), COMBO-17 J1146-01 (Wolf 2005),
and the faintest white dwarfs detected in the globular cluster
NGC~6397 (Richer et al. 2006; Hansen et al. 2007).\footnote{
In addition, there are some very cool white dwarfs
that probably show CIA but require additional data to confirm their
classsification:  LHS 2288 (Scholtz et al. 2005), CE 51 (Ruiz \&
Bergeron 2001), the companion to PSR J0751+1807 (Bassa et al. 2006),
and several candidates from SDSS (Kilic et al. 2006).}
Several have been individual and serendipitous discoveries.
However, the Sloan Digital Sky Survey (SDSS -- York et al. 2000;
Gunn et al. 1998; Stoughton et al. 2002; Gunn et al. 2006)
is suitable for systematically finding the coolest white dwarfs
through its accurate broadband photometry (Fukugita et al. 1996;
Hogg et al. 2001; Lupton et al. 2001; Smith et al. 2002; Ivez{\'i}c
et al. 2004; Tucker et al. 2006), spectroscopy of selected targets,
and astrometry (Pier et al. 2003; Munn et al. 2004).  The six very cool
white dwarfs discovered in the SDSS thus far have contributed half of
the known sample.  These increased numbers help address both of the
above goals of improving white dwarf models and understanding the
evolution of white dwarfs, but much larger samples are still needed.

In this paper, we present data for an additional seven
``ultracool'' white dwarfs discovered in the SDSS.
The wavelength coverage of SDSS data ($\lambda < 1 \mu$m)
allows discrimination of the coolest white dwarfs only when CIA
is sufficiently strong to affect their optical colors and spectra.
Therefore, mild CIA stars like WD0346 cannot be identified with
SDSS data alone, and are not included as new discoveries in this paper.
Following Gates et al., we use the term ``ultracool'' to refer to
white dwarfs with CIA detectable at wavelengths shorter than $1 \mu$m.

\section{Observations}

The ultracool white dwarfs in this paper have been found in
searches of SDSS spectra available to the SDSS Collaboration
through 2006.  The dataset is approximately coincident with the
SDSS DR6 data release \citep{ade07}, with spectroscopic data
covering 7425 deg$^2$ of the sky;  this paper increases
the sky coverage by about 70\% beyond that of the previous paper
by \citet{gat04}.  The seven stars in this paper were targetted
for spectroscopic observations under six different selection
categories described in Sec.~3 below.

Among the SDSS spectra of cool white dwarfs
with featureless DC spectra, there are a number that we have
classified as normal, not ultracool.  The distinction is not
always clear, because the SDSS spectra of faint stars ($g \sim 19$)
have poor signal/noise at the red end from the difficult subtraction
of the strong night sky emission, and because the $z$ magnitudes have
larger errors than the $gri$ magnitudes.  In this paper, we have relied
primarily on the $i-z$ color, requiring it to be significantly
bluer for the stars we classify as ultracool than for normal cool
white dwarfs.  Additional infrared observations are desirable
to further study where stars lie in the transition region from
normal cool white dwarfs, to white dwarfs with weak CIA detectable
only in the infrared, to the ultracool stars described in this paper.

Data for the seven new ultracool white dwarfs are given in Table 1,
and the SDSS spectra are shown in Figure 1.
Positions and magnitudes are taken from the SDSS database.
Proper motions, calculated as described by \cite{mun04},
are included in Table 1.  For two stars that are sufficiently bright
and unblended to provide reliable matches with objects in the
USNO-B catalog (Monet et al. 2003), proper motions are taken from
the SDSS database.  For five stars that are too faint
or are confused in USNO-B, we calculated new proper motions
using the Munn procedure, using the SDSS position plus those USNO-B
plate positions that did appear reliable.

Additional photometry in the Johnson/Cousins $BVRI$ system
has been obtained with the 40-inch telescope of the U.S. Naval Observatory
in Flagstaff.  These data were obtained in order to help confirm the
unusual colors of some of these stars caused by CIA, and to enable
comparisons with other stars not yet observed with SDSS filters.
Observations were made on eight nights during 2005-2007 with a
Tektronix CCD and standard filters.  Magnitudes and colors were
transformed to the standard system using Landolt (1992) standards.
The results are given in Table 2.

Only one of these seven new ultracool white dwarfs shows strong CIA --
in Fig. 1, SDSS J1251+44 has a maximum in its flux density $f_{\nu}$
at 5000\AA, bluer than any other ultracool white dwarf.  Its $B-V$
color of +0.28 is consistent with the very strong red and infrared
absorption by CIA indicated by the SDSS colors.  Its $griz$ and $BVI$
colors are more like an A star or a quasar than a cool white dwarf,
and create some doubt about its nature.  However, the reduced proper
motion given in Table 1 must be approximately correct\footnote{
By chance, SDSS J1251+44 lies on 14 POSS plates in the USNO
pixel database.  It is clearly detected on four plates,
marginally visible on four plates, and not visible on six plates
(all of the POSS-I E and POSS-II IV-N plates).
The proper motion given in Table 1 is somewhat uncertain depending
on whether the marginal detections are considered real and are
included or omitted from the solution.  Nevertheless, the definite
detections on POSS-II plates taken between 1988 and 1997, plus the
SDSS detection in 2003, show a clear westward motion.
},
and is much too large to allow classification as something other
than a white dwarf.  Furthermore, the lack of real features in the
(admittedly noisy) spectrum is not consistent with a horizontal branch
or blue straggler star.  Some new type of warm white dwarf
(perhaps magnetic?) might be possible, but the drop in flux at the
blue end of the spectrum and the red $u-g$ color are unlike any known
warm white dwarf.  A few DZ white dwarfs with strong absorption by
metals (Dufour et al. 2007) have spectra that are superficially similar,
but always show calcium and/or magnesium features that are absent
in SDSS J1251+44.  Therefore, we prefer a ``conventional''
classification of a white dwarf with strong CIA.

The remaining six stars have mild CIA, as shown by their red colors
and spectra.  The weakest CIA occurs in SDSS J0310$-$01, where the
$ugri$ colors are consistent with a normal cool white dwarf,
and only the $i-z$ color of $-$0.12 reveals its ultracool nature.
The result is that these seven white dwarfs have a surprisingly
large range of colors:  $g-r$ ranges from $-$0.21 to 1.43.

SDSS J2239+00 has a companion with a separation of 2.0$''$.
The companion star, SDSS J223954.07+001849.2,
has $griz$ magnitudes of 21.00, 19.94, 19.60, and 19.43.
(The $u$ magnitude of 24.1 is too faint and noisy to be useful.)
The pair is just visible on a UK Schmidt R plate taken in 1988,
with the separation and orientation appearing unchanged between
1988 and the imaging done by SDSS in 2001 and at USNO in 2005.
Therefore, the pair probably has common proper motion and is
a physical binary.  The separation is large enough that the
spectrum in Fig. 1 (taken with a 3$''$ fiber) should not have been
significantly contaminated by the companion, and, indeed, the
colors and spectrum appear to be mutually consistent.  The companion
is itself a cool white dwarf, as indicated by its colors and reduced
proper motion.  The $griz$ colors given above and the $BVRI$ colors
in Table 2 indicate it is a normal cool white dwarf.  The companion's
red $i-z$ color of +0.17 places it close to the red limit that
is observed for cool white dwarfs, and shows that its optical colors
have not yet begun to turn blue from CIA.  However, Vidrih et al.
(2007) have discovered from UKIDSS near-infrared photometry that the
companion is deficient in $K$ flux, indicating the onset of CIA.
Therefore, this is a pair of cool white dwarfs, where weak CIA
is affecting the companion's infrared flux and stronger CIA is
affecting the optical spectrum and colors of SDSS J2239+00.
It is the first such pair of CIA white dwarfs identified.
Vidrih et al. derive a distance of 55 pc, consistent with the
distance that we estimate below (Table 1) for the ultracool star.

\section{Discussion}

Color-color plots are shown in Figure 2 for all 15 ultracool
white dwarfs observed in SDSS;  these are the seven new stars in
this paper, the seven previous stars from Gates et al. (2004)
and Harris et al. (2001a), and the one previous star from Wolf (2005).
(This last star was present in the SDSS photometric database,
but was not designated as an SDSS spectroscopic target because
of its faint magnitude.)
The stars have been arbitrarily divided into two groups, one group
showing strong CIA in the spectra and colors (filled circles in
Fig. 3) and a peak in the flux density $f_{\nu}$ shortward of
7000\AA, and the other showing more mild CIA and a peak in
$f_{\nu}$ longward of 7000\AA.  This division is not meant to
imply that the groups are distinct, because there is probably a
continuous distribution of CIA strengths.  The two groups and symbols
only help to interpret the figures by highlighting stars which are
near the turnaround in color and stars which have cooled sufficiently
(for their individual atmospheric hydrogen/helium abundances)
to be well beyond the turnaround in color.

The new stars in this paper have filled in much of the gap in
colors between the previous ultracool stars and the warmer cool
white dwarfs that do not yet show CIA (shown by dots in Fig. 2).
The large variety of colors seen in Fig. 2 probably is caused
in part by a range of hydrogen/helium abundance ratios.
Colors of models from Bergeron with two abundance ratios are
shown in Fig. 2.  More recent models of pure hydrogen atmospheres
that include additional opacity from the far red wing of the
Lyman~$\alpha$ line \citep{ks06} are also shown, and extend the
range of model colors further to the red.  It can be seen that
the three models match the range of colors at which stars develop
mild CIA and turn off from the sequence of normal cool white dwarfs.
However, these models only qualitatively reproduce the range of
colors for stronger CIA.  A range of stellar masses, and
corresponding surface gravities, also may be a contributing factor.
Obviously, improved models are needed before they can be applied
to this full set of stars and be expected to yield quantitative
conclusions about temperature, abundance, and age.

The stars at the faint end of the white dwarf sequence in
NGC~6397 (Hansen et al. 2007) appear to have colors similar
to the mild CIA stars found in this paper.  In NGC~6397, the
faintest white dwarfs have a range in F606W$-$F814W of 0.4,
corresponding to a range in $V-I$ of 0.5 or a range in $g-i$
of 0.6.  Aside from SDSS J1238+35, which is somewhat redder,
this is exactly the range in $g-i$ seen in Fig. 2 for the
mild CIA stars.  However, the masses, ages, and origin of
the stars in this paper may be quite different from those in
NGC~6397, as is discussed next.

The diagram of reduced proper motion in the $g$ filter
(H$_g$ = $g$ + 5log($\mu$) + 5) is shown in Figure 3.  The star
with the largest value of H$_g$ is SDSS J1220+09, with H$_g = 24$,
and was noted by Gates et al. (2004) as being the best candidate
for an old halo white dwarf.  This situation has not changed --
the other stars, including the new ones reported in this paper,
have small enough proper motions that they are likely to have
a disk (or perhaps thick disk) motion and origin.
In Table 2 we include estimates of the distances and corresponding
tangential velocities for each of the new ultracool white dwarfs
based on assuming a wide range for the absolute magnitude
($M_V = 16.5 \pm 1.0$), following Salim et al. (2004)
and Gates et al. (2004).  Parallax measurements are essential for
reducing the uncertainties in these estimates and for understanding
the systematic properties (masses and disk/halo origins) of these
stars;  observations are in progress for measuring the parallaxes
of several of these stars with USNO and MDM telescopes.

The spectrum of SDSS J0310$-$01 exhibits narrow H$_{\alpha}$ emission
(and  probably H$_{\beta}$) which suggests the presence of an
unseen companion.  Additional data at optical and/or IR wavelengths
are needed to confirm this possibility and determine the nature
of the companion, and will be reported in a future publication.
Judging from Figure 2 of Dahn et al. (2002) giving M$_I$ vs.
spectral type, an L0 companion would contribute M$_I \sim 15$--15.5,
comparable to the likely absolute $I$ mag of the white dwarf;
this limit suggests that the companion should be at least early L.
Admittedly, a rising contribution from the companion could be partially
cancelled by the CIA decline the white dwarf's energy distribution might
have.  Since such objects are generally not chromospherically active,
and there is no significant ultraviolet radiation field, the narrow
H$_{\alpha}$ emission suggests the possibility of a close binary
with rapid rotation perhaps synchronous with the orbital period.
We have examined the individual exposures (13 exposures for this star)
used in the SDSS spectrum in Fig. 1, but they are noisy enough to
mask any changes in the emission that might be occurring.
Observations to search for periodic radial velocity variations should
be obtained.

It is premature to add these new ultracool white dwarfs to
any analysis of the space density and luminosity function
of white dwarfs for two reasons:  we do not yet have models
to fit the spectra adequately to give accurate temperatures
and H/He abundances, and we do not yet have distances to get
luminosites, masses, and ages.  Nevertheless, we can make
a few comments about the effects these new stars might have
on our understanding of the WDLF.  First, only one of these
stars (SDSS J1632+24) is bright enough to be discovered
using Palomar Sky Survey plates, and its proper motion is
too small to enter into the sample used for the WDLF by
Liebert et al. (1988), or into the expanded sample with a
lower proper motion limit being used by Harris et al. (2001b).
Second, all of these new stars would have been detected in
the deeper white dwarf sample by Knox et al. (1999) and
contribute to their LF.  However, their search covered only
25~deg$^2$, compared to the 7425 deg$^2$ in the SDSS DR6 area.
Therefore, even if the SDSS spectroscopic sample is incomplete
by a substantial factor for these stars with mild CIA, it is
unlikely that even one such star is actually present in their
sample.  Third, none of these stars have $g < 19.5$, so none
would have entered the sample used by Harris et al. (2006)
to measure the WDLF with SDSS imaging data.
Fourth, the one binary among these stars is too faint to be
discovered on Sky Survey plates, so would not enter the
sample of binaries used for the WDLF by Oswalt et al. (1996).
Fifth, only one of these stars (SDSS J1632+24) has a proper
motion large enough to enter the sample of halo WD candidates
found by Oppenheimer et al. (2001), and their analysis would
have assigned it a distance of $\sim$50~pc and $V_{\rm tan}
\sim 80$ km~s$^{-1}$, so it would have been classified a
disk star.

These comparisons suggest that none of the determinations of
the WDLF up to now are affected by ultracool white dwarfs
of the type found in this paper.  In order to measure the LF
that {\it does} include them, it is necessary to go to fainter
magnitudes and/or larger areas of sky in order to sample a
larger volume of space.  One example is the 90-Prime survey
(Liebert et al. 2007), a proper-motion followup to SDSS that
is reaching about 2 mag fainter than the photographic surveys
and will take advantage of the deep limiting magnitude of SDSS.
A second example that may help determine the density of
ultracool white dwarfs is the Sloan Extension
for Galactic Understanding and Exploration (SEGUE),
one of three surveys that comprise the second phase of SDSS.  
Of the seven new discoveries reported in this paper, four were
targeted for spectra as part of the SDSS main survey and three
were targeted for spectra taken for SEGUE.\footnote{
SDSS J1238+35 was targeted by the QSO target selection category,
SDSS J1251+44 by SERENDIPITY\_DISTANT, and SDSS J1452+45 and
SDSS J1632+24 by STAR\_WHITE\_DWARF (see Gates et al. 2004;
Stoughton et al. 2002; Richards et al. 2002).
SDSS J0310$-$01 and SDSS J2239+00 were targeted as cool white dwarf
candidates by SEGUE, and SDSS J0146+14 was found serendipitously in
early SEGUE spectra through preliminary target selection for red dwarfs.
}
SEGUE includes a target selection algorithm based on proper motion
that is designed to include cool white dwarfs of all types, and should
allow a complete sample to be identified of these ultracool white
dwarfs with mild CIA.  The SEGUE survey ultimately
will cover 3500 square degrees in imaging, and obtain spectra of
stars along 200 lines of sight.  Ten spectroscopic fibers
per plate pair are being assigned to cool white dwarf candidates,   
for a total of roughly 2000 spectra.  The density of 0.0014~deg$^{-2}$
stars with strong CIA with $i<20.2$ \citep{gat04} indicates that only
two strong-CIA stars will be found in SEGUE, and probably two-four
additional weak-CIA stars.  However, the true density may be higher
(Wolf 2005).  Following this work, the PanSTARRS survey, and
ultimately LSST, have the potential to greatly improve the complete
samples that are needed for the WDLF.

\acknowledgments

P. Bergeron, P. Kowalski, and D. Saumon kindly calculated and
made available SDSS colors of their model atmospheres.
This research has made use of the USNOFS Image and Catalogue Archive
operated by the United States Naval Observatory, Flagstaff Station
(http://www.nofs.navy.mil/data/fchpix).

Funding for the creation and distribution of the SDSS Archive has
been provided by the Alfred P. Sloan Foundation, the Participating
Institutions, the National Aeronautics and Space Administration,
the National Science Foundation, the U.S. Department of Energy, the
Japanese Monbukagakusho, and the Max Planck Society. The SDSS Web
site is http://www.sdss.org/.

The SDSS is managed by the Astrophysical Research Consortium (ARC)
for the Participating Institutions. The Participating Institutions
are The University of Chicago, Fermilab, the Institute for
Advanced Study, the Japan Participation Group, The Johns Hopkins
University, the Korean Scientist Group, Los Alamos National Laboratory,
the Max-Planck-Institute for Astronomy (MPIA),
the Max-Planck-Institute for Astrophysics (MPA),
New Mexico State University, University of Pittsburgh,
Princeton University, the United States Naval Observatory,
and the University of Washington.

\clearpage

\begin{deluxetable}{lccccccc}
\rotate
\tabletypesize{\scriptsize}
\tablecolumns{8}
\tablewidth{0pt}
\tablecaption{New Ultracool White Dwarfs}
\tablehead{
\colhead{Short Name} &
\colhead{J0146+14} &
\colhead{J0310$-$01} &
\colhead{J1238+35} &
\colhead{J1251+44} &
\colhead{J1452+45} &
\colhead{J1632+24$^2$} &
\colhead{J2239+00} }
\startdata
R.A.$^1$  &01 46 29.01 &03 10 49.53 &12 38 12.85 &12 51 06.12
      &14 52 39.00 &16 32 42.23 &22 39 54.12 \\
Dec.$^1$  &+14 04 38.2 &$-$01 10 35.3 &+35 02 49.1 &+44 03 03.1
      &+45 22 38.3 &+24 26 55.2 &+00 18 47.3 \\
Epoch$^1$ &1999.7818&2002.6779&2004.2905&2003.2310
      &2003.4059&2003.4693&2001.7870 \\
$\mu$          (mas yr$^{-1}$)& 255&   88&   180&   170&
        93&  340&   98 \\
$\mu_{\alpha}$ (mas yr$^{-1}$)& 252&$-$36&$-$130&$-$167&
     $-$54& $-$10&   7 \\
$\mu_{\delta}$ (mas yr$^{-1}$)&  38&$-$80&$-$124&    30&
        76&$-$340&  98 \\
$u$ &21.19 &22.37 &24.74 &21.43 &21.55 &21.33 &21.46 \\
$g$ &19.97 &20.93 &21.73 &20.18 &20.06 &19.60 &20.15 \\
$r$ &19.38 &20.18 &20.30 &20.39 &19.39 &18.72 &19.52 \\
$i$ &19.26 &19.87 &19.86 &20.69 &19.31 &18.51 &19.49 \\
$z$ &19.70 &19.99 &20.32 &20.90 &19.38 &18.47 &20.09 \\
H$_g$$^3$ & 22.00& 20.64& 23.00& 21.33& 19.91& 22.26& 20.11 \\
Distance (pc)&26-67&40-99&48-121&36-90&27-68&21-52&28-72 \\
v$_{\rm tan}$(km s$^{-1}$)&32-80&17-42&41-103&29-73&12-30&33-84&13-33 \\
MJD/Plate/Fiber$^4$&53262-1899-493 &53386-2068-054 &53431-2020-417
   &53063-1373-176 &53466-1675-014 &53226-1573-635 &53261-1901-400 \\
\enddata
\tablenotetext{1}{Coordinates are given for equinox J2000.0 at the
observed epoch.
Units of right ascension are hours, minutes, and seconds,
and units of declination are degrees, arcminutes, and arcseconds.}
\tablenotetext{2}{SDSS J1632+24 previously found as
   LP386$-$28 (Luyten 1979) = WD1630+245 (McCook \& Sion 1999).}
\tablenotetext{3}{Reduced proper motion (H$_g$ = $g$ + 5$log \mu$ + 5).}
\tablenotetext{4}{SDSS spectra: MJD of observation--plate
number--fiber number.}
\end{deluxetable}

\clearpage

\begin{deluxetable}{lccccc}
\tablecolumns{6}
\tablewidth{0pt}
\tablecaption{$BVRI$ Photometry}
\tablehead{
\colhead{Star} &
\colhead{$V$} &
\colhead{$B-V$} &
\colhead{$V-R$} &
\colhead{$V-I$} &
\colhead{N$_{\rm obs}$} }
\startdata
SDSS J0146+14& 19.74$\pm$0.07& \dots        &$\phn$ 0.58$\pm$0.09&$\phn$ 0.85$\pm$0.09&1 \\
SDSS J0854+35& 19.80$\pm$0.03& \dots        &$\phn$ 0.67$\pm$0.04&$\phn$ 1.15$\pm$0.04&1 \\
SDSS J0947+44A$^1$& 19.12$\pm$0.02& 0.76$\pm$0.02&$\phn$ 0.40$\pm$0.02&$\phn$ 0.45$\pm$0.03&2 \\
SDSS J0947+44B$^2$& 19.09$\pm$0.02& 0.93$\pm$0.03&$\phn$ 0.52$\pm$0.02&$\phn$ 1.04$\pm$0.02&2 \\
SDSS J1001+39& 19.68$\pm$0.02& 0.74$\pm$0.04&$\phn$ 0.22$\pm$0.03&$-$0.08$\pm$0.05&2 \\
SDSS J1220+09& 19.67$\pm$0.03& 1.23$\pm$0.07&$\phn$ 0.48$\pm$0.04&$\phn$ 0.43$\pm$0.09&3 \\
SDSS J1251+44& 20.18$\pm$0.03& 0.28$\pm$0.05&$-$0.15$\pm$0.05&$-$0.20$\pm$0.07&4 \\
SDSS J1403+45& 18.81$\pm$0.02& 0.56$\pm$0.04&$-$0.15$\pm$0.03&$-$0.54$\pm$0.05&2 \\
SDSS J1452+45& 19.68$\pm$0.03& 0.77$\pm$0.08& \dots        &$\phn$ 0.60$\pm$0.08&1 \\
SDSS J1632+24& 19.09$\pm$0.03& 1.04$\pm$0.04&$\phn$ 0.51$\pm$0.03&$\phn$ 0.95$\pm$0.03&2 \\
SDSS J2239+00A$^1$&19.78$\pm$0.04& 0.70$\pm$0.07&$\phn$ 0.38$\pm$0.07&$\phn$ 0.61$\pm$0.05&2 \\
SDSS J2239+00B$^2$&20.23$\pm$0.05& 1.11$\pm$0.13&$\phn$ 0.75$\pm$0.07&$\phn$ 1.18$\pm$0.06&2 \\
\enddata
\tablenotetext{1}{A -- the ultracool white dwarf.}
\tablenotetext{2}{B -- the common-proper-motion companion white dwarf.}
\end{deluxetable}

\clearpage
\thispagestyle{empty}
\begin{figure}
\vspace*{-10mm}
\plotone{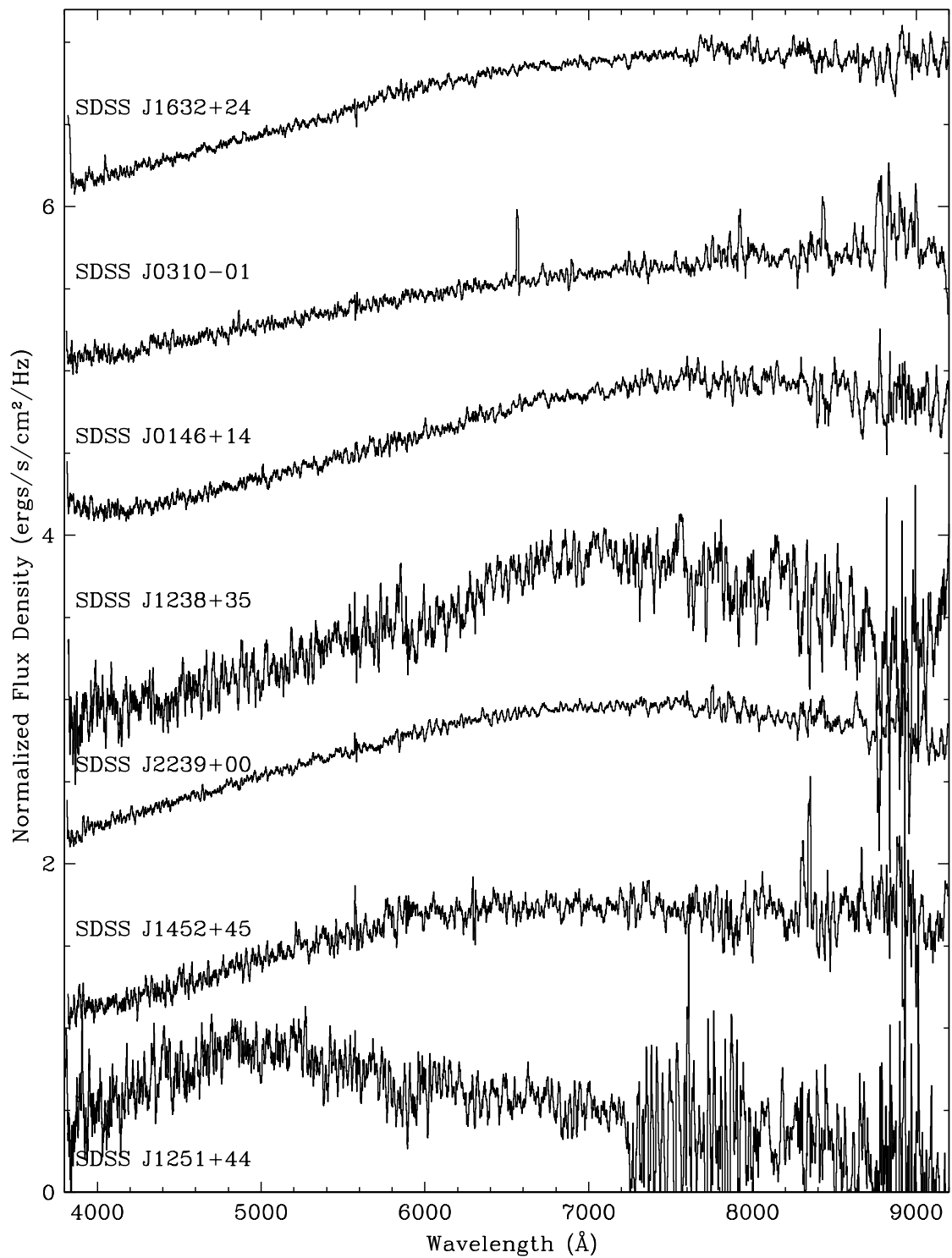}
\caption{SDSS spectra for the seven new white dwarfs in this paper.
Spectra have been smoothed by nine pixels to a resolution of 800,
and are normalized and offset vertically by one unit for display.
Compare with Fig. 2 of Gates (2004) for SDSS spectra of previously
observed cool white dwarfs.
\label{Fig.1}}
\end{figure}

\clearpage
\begin{figure}
\plotone{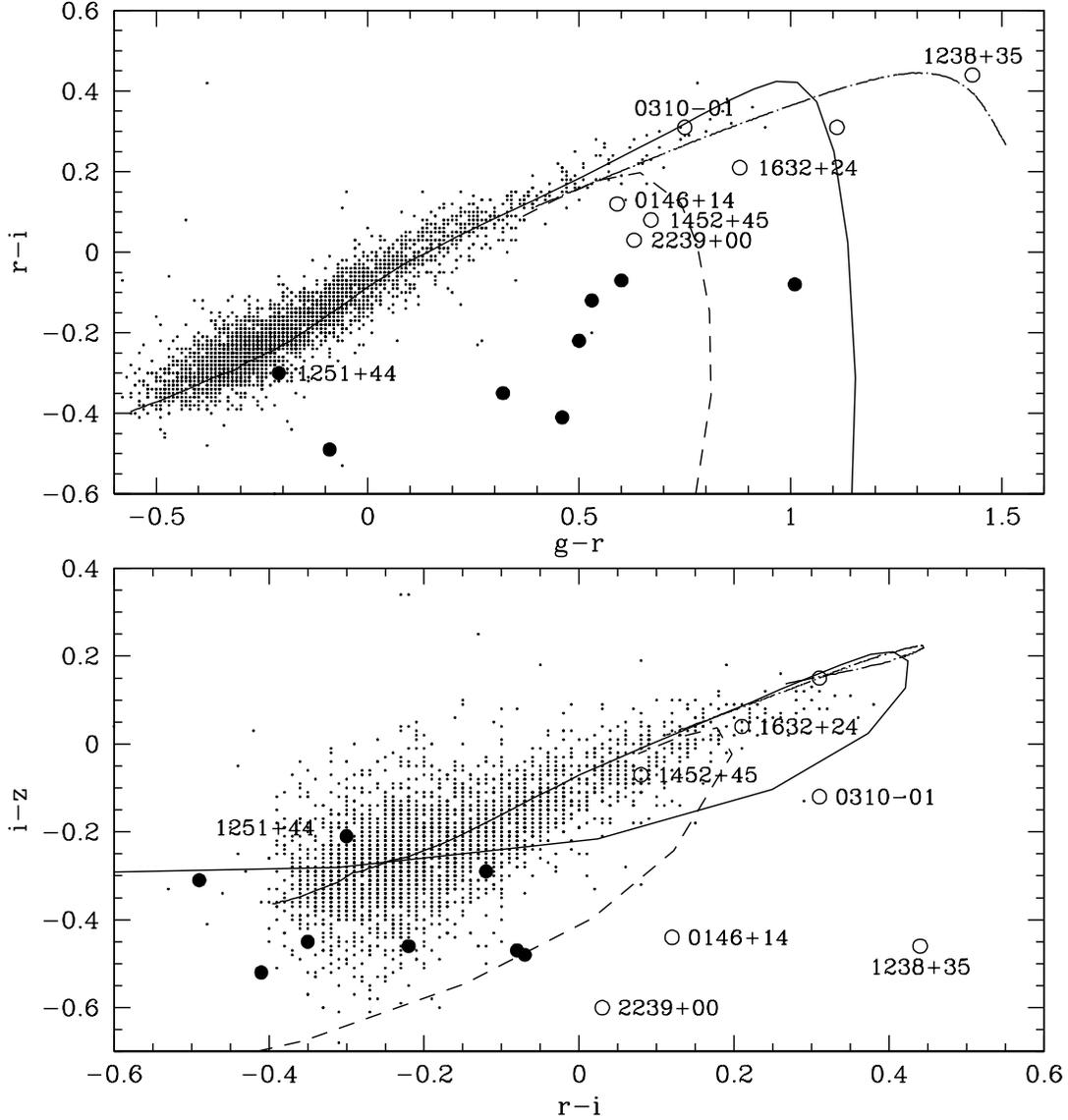}
\caption{Color-color plots for ultracool white dwarfs in SDSS colors.
Ultracool white dwarfs observed by SDSS are shown as filled circles
for strong-CIA stars and open circles for mild-CIA stars.
New stars identified in this paper are labeled.
The proper-motion-selected white dwarfs used for the luminosity
function by Harris et al. (2006) are shown as dots for comparison.
The curves show models for white dwarfs with log $ g = 8$ with
atmospheres of pure hydrogen (Bergeron et al. 1995b; solid line),
pure hydrogen (Kowalski \& Saumon 2006; dot-dashed line), and
He/H = 10$^5$ (Bergeron \& Leggett 2002; short dashed line).
\label{Fig.2}}
\end{figure}

\clearpage
\begin{figure}
\plotone{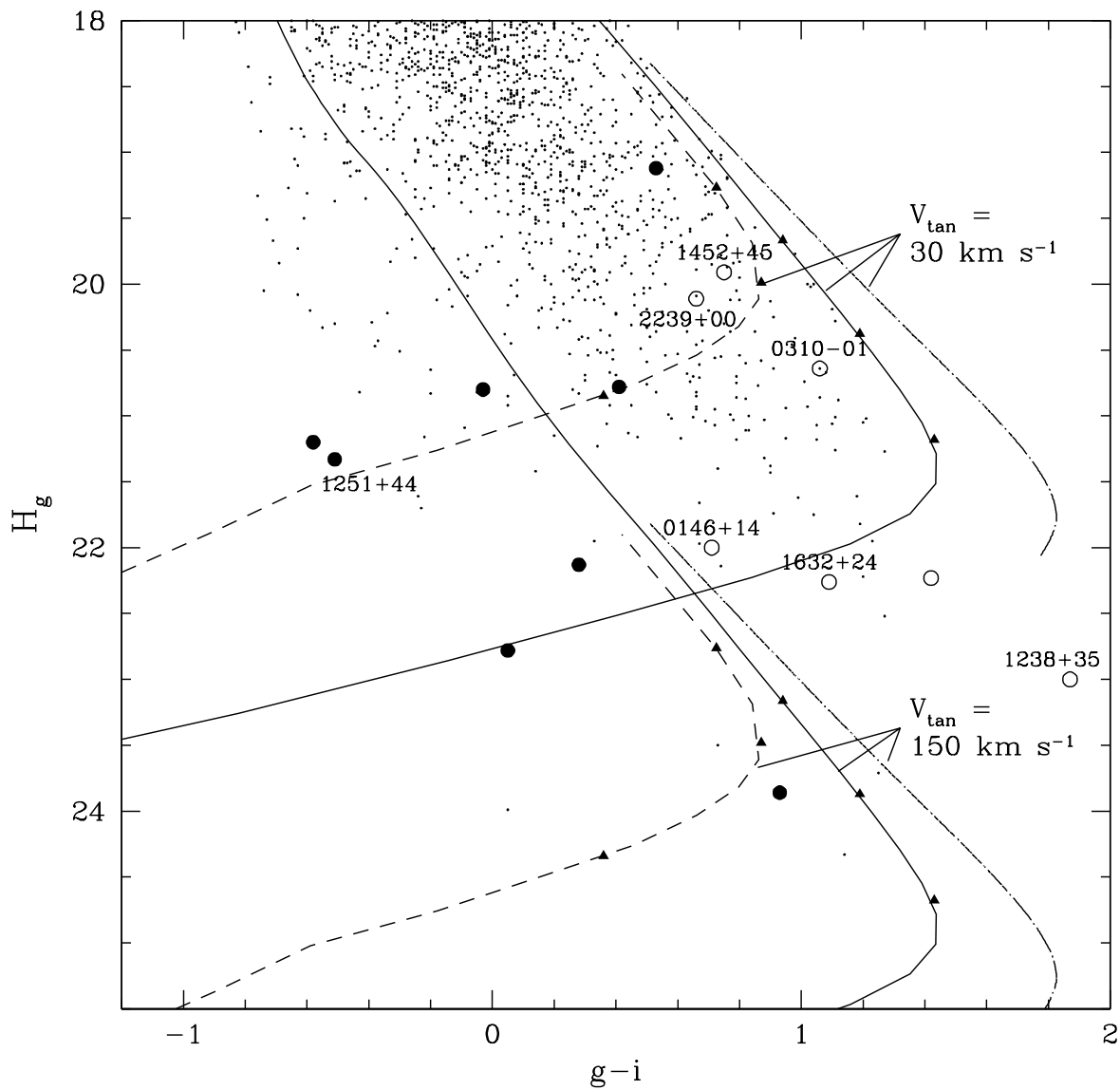}
\caption{Reduced proper motion for ultracool white dwarfs observed
by SDSS.  The symbols and curves are the same as in Fig. 2.
The model curves are plotted for two different assumed values
of the tangential velocity, again assuming log $ g = 8$.
On each curve showing Bergeron models (solid line for hydrogen,
short dashed line for mixed composition), three triangles indicate
ages of 6, 8, and 10 Gyr (top to bottom).
\label{Fig.3}}
\end{figure}

\end{document}